\documentclass{article}

\usepackage{arxiv}

\usepackage[utf8]{inputenc} % allow utf-8 input
\usepackage[T1]{fontenc}    % use 8-bit T1 fonts
\usepackage[hidelinks]{hyperref}       % hyperlinks
\usepackage{url}            % simple URL typesetting
\usepackage{booktabs}       % professional-quality tables
\usepackage{amsfonts}       % blackboard math symbols
\usepackage{nicefrac}       % compact symbols for 1/2, etc.
\usepackage{fancyhdr}       % header
\usepackage{graphicx}       % graphics
\usepackage{amsmath}
\usepackage{setspace}
\usepackage{xcolor}
\usepackage{caption, subcaption}
\usepackage{algpseudocode}
\usepackage{algorithm}
\DeclareMathOperator*{\argmin}{argmin}

\graphicspath{{media/}}     % organize your images and other figures under media/ folder

\title{Exploring the synergistic potential of quantum annealing and gate model computing for portfolio optimization}

\author{
  Naman Jain \\
  TCS Research, India \\
  \texttt{j.naman4@tcs.com} \\
   \And
 M Girish Chandra\\
  TCS Research, India \\
  \texttt{m.gchandra@tcs.com} \\
}

\begin{document}
\onehalfspacing
\maketitle

\begin{abstract}
Portfolio optimization is one of the most studied problems for demonstrating the near-term applications of quantum computing. However, large-scale problems cannot be solved on today's quantum hardware. In this work, we extend upon a study to use the best of both quantum annealing and gate-based quantum computing systems to enable solving large-scale optimization problems efficiently on the available hardware. The existing work uses a method called Large System Sampling Approximation (LSSA) that involves dividing the large problem into several smaller problems and then combining the multiple solutions to approximate the solution to the original problem. This paper introduces a novel technique to modify the sampling step of LSSA. We divide the portfolio optimization problem into sub-systems of smaller sizes by selecting a diverse set of assets that act as representatives of the entire market and capture the highest correlations among assets. We conduct tests on real-world stock data from the Indian stock market on up to 64 assets. Our experimentation shows that the hybrid approach performs at par with the traditional classical optimization methods with a good approximation ratio.  We also demonstrate the effectiveness of our approach on a range of portfolio optimization problems of different sizes. We present the effects of different parameters on the proposed method and compare its performance with the earlier work. Our findings suggest that hybrid annealer-gate quantum computing can be a valuable tool for portfolio managers seeking to optimize their investment portfolios in the near future.

\end{abstract}

% keywords can be removed

\section{Introduction}

Portfolio optimization is the problem of selecting the best distribution of assets that optimises a particular objective function \cite{Markowitz1952}. Usually, this objective function attempts to minimize the risk and maximize the expected returns. It is a complicated problem that has been the subject of extensive research in finance, computer science and mathematics. The complexity of this problem arises from several factors such as the large number of assets available for investment, the dynamic nature of market conditions and various constraints that must be considered. Optimizing a portfolio involves balancing the trade-off between the expected returns and the risk by selecting the optimal combination of assets. This is a challenging task because the risks and returns associated with different assets are often interdependent and continuously changing. A well-optimized portfolio can provide significant benefits, including improved returns, reduced risks and increased diversification to achieve an efficient portfolio that is tailored to the investment goals and the risk tolerance of an investor. In practice, it is well-known that the portfolio optimization problem is generally computationally intractable. 

Classical optimization methods such as mean-variance optimization, and Monte Carlo simulations, have been widely used in the finance industry. However, these methods have limitations when dealing with large-scale problems, non-linear constraints and non-convex objective functions. Quantum computing methods, viz. quantum annealing \cite{robustness_of_aqc, Johnson2011} and gate-based quantum computing can potentially solve complex optimization problems more efficiently than classical methods and may provide better solutions for practical problems with many variables and constraints. It has been understood recently that quantum and quantum-inspired computing can help in tasks such as Monte-Carlo simulations \cite{option_pricing, quantum_finance_monte_carlo_pricing} and combinatorial optimization problems \cite{qaoa, robustness_of_aqc, hybrid_qc_comb} in many industries. Applications of quantum optimization to real-world problems have been demonstrated for portfolio optimizations \cite{portfolio40, portfolio_investment_bands}, detection of arbitrage cycles \cite{finding_arbitrage}, Travelling salesman problems \cite{Salehi_2022} and many more. To solve problems on quantum computers, an appropriate mapping of the original problem into a quantum-solvable one is required. A commonly studied class of combinatorial optimization problems are Quadratic Unconstrained Binary Optimization (QUBO) problems. There are multiple approaches to solving QUBO problems on a quantum device. 

Quantum annealing is a meta-heuristic utilized by adiabatic quantum computers. The QUBO problem is mapped to an Ising Hamiltonian whose ground state solution is related to the solution of the original problem \cite{lucas_ising_formultions}. It resembles simulated annealing and is applied to determine near-optimal solutions to QUBO problems. There are also variational quantum optimization heuristics such as the Variational Quantum Eigensolver (VQE) and Quantum Approximate Optimization Algorithm (QAOA) \cite{qaoa} to solve QUBO problems. VQE operates by using a set of parameterised gates to construct an ansatz (trial) state and uses a classical optimizer to optimise the parameters that best approximate the ground state of the problem Hamiltonian. QAOA on the other hand alternatively applies a series of operators which in infinite depth limit, would recover the adiabatic evolution and converge to the optimal solution. These heuristics are designed for near-term, noisy quantum machines without performance guarantees. 

Although several studies show remarkable results in portfolio optimization using the above-described common methods, these approaches require an \textit{N}-qubit quantum computer to solve the problem with \textit{N} assets. This limits their usage to small problems as the largest available gate-based quantum computer to date is the Osprey processor from IBM with 433 physical qubits \cite{IBMQuantum} and the largest annealer is the DWave 5000-qubit system on the Pegasus chip with 15 connections per qubit \cite{Dwave}. There have been proposals to mitigate this issue by dividing the original problem into smaller sub-problems and then re-combining the solutions to approximate the full problem solution. 

In this paper, we extend upon a study by Liu et al. \cite{liu2022hybrid}, to utilize both quantum annealing and gate-based quantum computing for the case of portfolio optimization. Their proposition is to divide a large Ising problem into smaller sub-systems, solve the sub-systems on available quantum hardware and then recombine the solutions. They propose a hybrid structure that solves the sub-system problem either by annealer or gate-based chips and then combines the solutions with amplitudes optimized using VQE on a gate-based quantum computer. This technique allows to solve much larger problems efficiently on the available hardware. Here we introduce a novel technique to sample sub-systems by selecting assets such that they form groups that capture the maximum dependencies between variables. To do this, we build a market graph of the assets and select a set of diverse assets among them using the Maximum Independent Set (MIS) of the graph. The assets in the MIS represent the entire market intuitively and so are used to build the sub-systems. 

We investigate a QUBO formulation of a simplified version of the portfolio optimization problem and test the proposed method on real-world stock datasets. In the process, we also compare the performance of the method against previous techniques for varying numbers of assets and changing different parameters. Overall, our study contributes to the growing body of research on quantum computing in finance and is a step towards practical implications for investors and asset managers.

The rest of the paper is organized as follows. In Section \ref{portfolio_formulation}, we introduce the portfolio optimization problem and its QUBO formulation. We also discuss traditional classical algorithms and their limitations. In Section \ref{methdology}, we briefly discuss the previous research - LSSA \cite{liu2022hybrid}, followed by the proposal of our method. In Section \ref{implementation}, we document the experimental parameters and implementation details. Further, in Section \ref{results_and_discussion}, we present our findings on actual stock data and also provide a comparison of the performance of different methods. Finally, in Section \ref{conlusion}, we conclude our study and provide directions for future research.

\section{Portfolio Optimization} \label{portfolio_formulation}

Portfolio optimization is aimed at creating an investment portfolio that maximizes returns while minimizing risk. Here, we consider the portfolio optimization problem expressed in a QUBO formulation \cite{portfawn, Mugel_2022}

\begin{equation} 
    H = - \mu^{T}\omega + \gamma \omega^{T}\Sigma \omega  \label{pf_form}
\end{equation}

where $\omega$ is an \textit{N}-dimensional vector of binary decision variables, $\mu$ is the vector of expected returns and $\Sigma$ is the covariance matrix of the returns on the assets in the portfolio. The term $\mu^{T}\omega$ represents the expected return on the portfolio and the term $\gamma \omega^{T}\Sigma \omega$ denotes the variance of portfolio return. $\gamma \ge 0$ is the risk-aversion factor and indicates the risk appetite of the investor. In this relaxed formulation, we assume that only long positions are possible. We suppose that the total budget is equally distributed among the selected assets and that the risk is estimated as the volatility, which is the square root of the portfolio variance. We also assume a static nature and do not consider the changing market conditions or investor preferences.

Classical algorithms for solving portfolio optimization problems, include Markowitz mean-variance optimization, which aims to maximize expected returns while minimizing variance, and the Capital Asset Pricing Model (CAPM), which focuses on estimating the expected returns of assets based on their systematic risks. Other popular approaches include the Sharpe ratio and the Black-Litterman model, which incorporate additional factors such as transaction costs and investor preferences. While these methods have been widely used, they have certain limitations which can make them infeasible for large portfolios. There is an exponential increase in the number of computations required as the number of assets in a portfolio increases. These methods can also get stuck in local optima, leading to sub-optimal solutions. Quantum optimization methods can likely overcome these limitations to deliver better portfolio allocations and potentially higher returns. 

\section{Methodology} \label{methdology}

\subsection{Large System Sampling Approximation } \label{method_lssa}

LSSA \cite{liu2022hybrid} divides a full Ising problem of \textit{N} variables into smaller $N_s$ sub-system problems each of size $N_g$ variables ($N_g \leq N$). The sub-systems are solved independently considering the original problem Hamiltonian either on annealing or gate-based quantum chip. The solutions of these sub-systems are then recombined by optimizing the amplitude contributions of each of them by using a VQE on a gate-based quantum computer. The full problem solution is a statistical mixture of sub-problem solutions. The complete mathematical description of this procedure is described in the following paragraphs.

An Ising problem of the below form is considered 

\begin{equation} 
    H = \sum_{i,j=1}^{N}J_{i,j} z_i z_j + \sum_{i=1}^{N}h_i z_i \label{ising}
\end{equation}

where $z_i$ are the spin variables, $z_i \in \{-1, +1\}$, \textit{N} is the total number of variables, and $h_i$ and $J_{i,j}$ correspond to the bias and coupling strength of the spin system respectively. Ising and QUBO formulations are interchangeable by means of the transformation $z_i = 2x_i - 1$ where $x_i \in \{0, 1\}$ is the binary variable and $z_i \in \{-1, +1\}$ is the spin variable. 

The sub-systems are created by randomly sampling $N_g \leq N$ sites $N_s$ times. The sampling procedure guarantees that all variables are picked at least $\lfloor (N_s \times N_g)/N \rfloor$ times.  The new sub-Hamiltonians remain in the same form as \eqref{ising}, only containing the variables selected in the corresponding sub-systems. 

The eigenvalue problem of the reduced Hamiltonian is solved and the corresponding ground state is labelled as $| GS_{sub}^{(i)} \rangle$, where \textit{i} represents the $i^{th}$ sub-system. This gives a partial spin-configuration of the full system, that is, a vector of length \textit{N} with $N_g$ non-empty sites. Each element in this vector is either $+1, -1 \; or \; 0$, corresponding to selecting, rejecting and variable not being present in that sub-system conditions. At the end of this procedure, there are $N_s$ such vectors that are the ground state solutions of the $N_s$ sub-systems.

The $N_s$ sub-systems are then aggregated as a weighted sum of the sub-system ground states,
\begin{equation} 
    |\mathcal{S}^{wc} \rangle = \sum_{i=1}^{N_s} C^{(i)}| GS_{sub}^{(i)} \rangle \label{Ns_agg}
\end{equation}

Further, the ground state configuration of the full system is approximated by the sign of each variable in $|\mathcal{S}^{wc} \rangle$.
\begin{equation} 
    |sign(S^{wc} \rangle \approx |GS_{full} \rangle  \label{sign_swc}
\end{equation}

It is expected that the above approximation is accurate when the sub-system size approaches the full system size, i.e., as $N_g \longrightarrow N$.

To determine the coefficients $C^{(i)}$ in \eqref{Ns_agg}, VQE is employed to minimize the expectation value of the full-system Hamiltonian \textit{H} and determine the ground state of the full-system. The $N_s$ coefficients are encoded into the amplitudes of the quantum state, using $N_{gb} = \lceil log_{2} N_s \rceil$ qubits of a gate-based system, and the cost function is defined as

\begin{equation} 
    \text{Cost}(\textbf{C}(\overrightarrow{\theta})) = \langle sign(S^{wc}) | H | sign(S^{wc}) \rangle \label{vqe_cost}
\end{equation}

where $\overrightarrow{\theta}$ represents a set of tunable parameters of the VQE.

LSSA enables solving large problems on the available quantum hardware. The sub-systems could be solved by an annealer which usually has greater qubits than the gate-based systems. This hybrid structure uses the best of both kinds of quantum systems to solve large Ising problems efficiently. We refer the reader to \cite{liu2022hybrid} for a complete description and analysis of LSSA.

One of the possible limitations of LSSA is the number of samples $N_s$. Ideally, to ensure that quadratic interactions between all the variables are captured, $N_s = \binom{N}{N_g}$ samples are required. This number grows combinatorially and hence a very delicate trade-off between $N_g$ and $N_s$ is needed to ensure acceptable performance and the problem being accommodatable on the available quantum machines. However, randomly selecting the sub-systems would require a very large number of samples and consequently many expensive calls to the quantum chips. Moreover, choosing fewer samples might lead to sub-optimal results as the strongest couplings between the variables might be disregarded in the random process. Thus a better sampling method is needed to have quality small number of samples.

\subsection{Proposed Method}

\subsubsection{Level 1 LSSA with MIS} \label{lvl1}

\begin{figure}
  \centering
  \includegraphics[width=\linewidth]{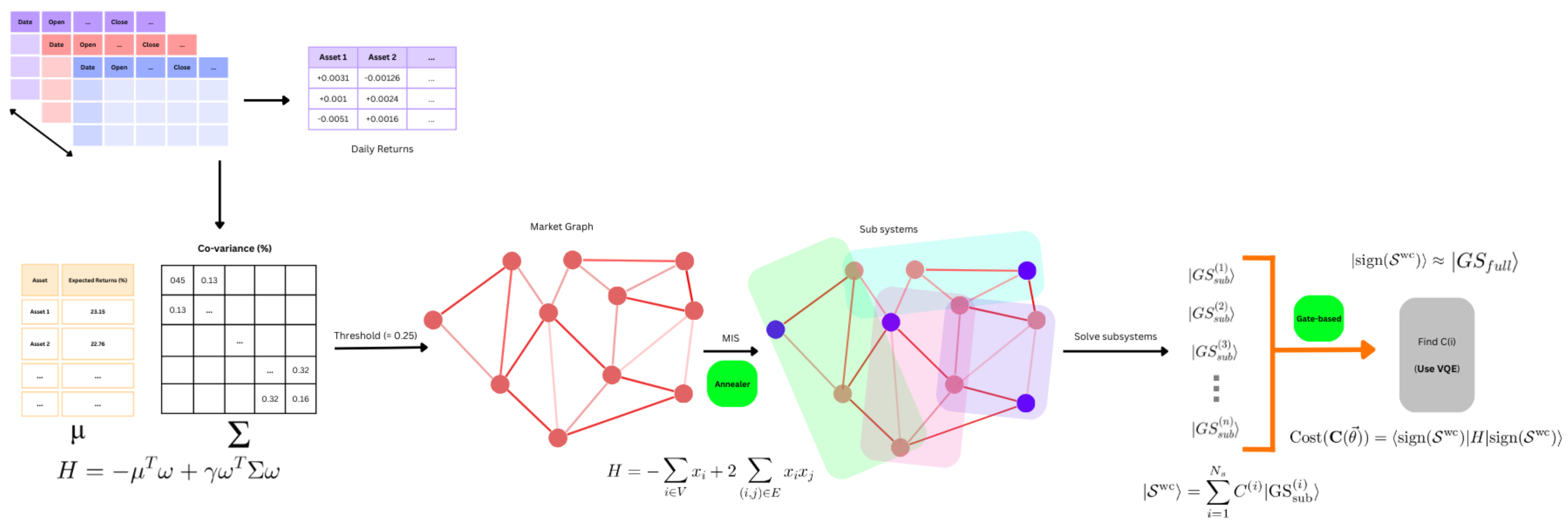}
  \caption{Flow of the proposed method. The first step is obtaining the expected returns vector and the covariance (and correlation) matrix from the data. The correlation matrix is thresholded to create the market graph. The MIS of the graph gives assets used to create sub-systems. The MIS problem may be further decomposed as described in Section \ref{lvl2}. The sub-systems are then solved and combined following the LSSA methodology.}
  \label{fig:flow}
\end{figure}

The process of selecting the sub-systems in LSSA at random is not perfect. Important and significant couplings between the variables might be omitted simply due to the nature of the sampling. To better select the sub-systems, we propose a coupling-dependent sampling methodology based on the diversification of the portfolio to select assets that are representative of the entire market and capture the highest couplings between all the variables. The complete description of the proposed method is presented in the following paragraphs.

Assets in the market are usually correlated with each other. The strength of the correlation between two assets is an indication of the risk of investing in both assets. Hence, an investment in negatively correlated assets is generally rewarding, while an investment in strongly positively correlated assets is risky. Using the absolute value of correlation as a selection criterion, a graph $G = (V, E)$ of assets is constructed with the asset symbol set as the vertices, $V$, and the edge connection, $E$, determined by the selection criteria. Therefore, an edge is drawn between two vertices if the corresponding pair of assets have an absolute correlation value above some static threshold value $\alpha$. The MIS of this market graph produces a diversified portfolio \cite[Chapter~4]{finance_book}, but only considering the correlation measure. The assets present in the MIS are then used to generate the sub-systems. A thorough explanation of the mechanism to accomplish this task is provided subsequently.

MIS is a maximal set of vertices in a graph such that no two of them are adjacent. It is a classic NP-Hard graph problem \cite{mis_np_hard} that has been studied extensively. There is also a well-known QUBO formulation for MIS that facilitates the use of quantum optimization techniques to address the problem.

\begin{equation}
    H = -\sum_{i \in V} x_i + 2 \sum_{(i,j) \in E} x_i x_j \label{mis_qubo}
\end{equation}

The optimization routine to achieve the ground state can be either annealing or gate-based quantum systems. The formulation in \eqref{mis_qubo} requires \textit{N} logical qubits for an MIS problem of \textit{N} variables. It might seem that there is no decrease whatsoever in the number of qubits. To handle this,  we add a second-level division to address the MIS problem. The mechanism to achieve that is explained in Section \ref{lvl2}

The assets selected in the MIS are indicated by the ground state of the Hamiltonian in \eqref{mis_qubo}. These assets act as placeholders for dividing the full problem.  The market graph contains many assets connected to each of those present in the MIS. We propose to make sub-systems out of each of these groups. Therefore, if there are $N_m$ assets in the MIS, then we create $N_s = N_m$ sub-systems such that each of those contains a group of assets connected to the corresponding placeholder asset in the MIS. However, for practical purposes, we limit the sub-system size to $N_g$ (dependent on the available quantum hardware). Eventually, we have $N_s$ sub-systems each containing a maximum of $N_g$ most correlated assets connected to each of the placeholder assets. After the creation of the sub-systems, the LSSA methodology shown in Section \ref{method_lssa} is followed. The sub-systems are independently solved and the solutions are recombined using a VQE with a circuit structure similar to \cite{liu2022hybrid}. The entire process is shown in Fig. \ref{fig:flow}  for a better understanding.

Since correlation is not a transitive property, it is possible for an asset to be present in many sub-systems, even though their placeholder assets are not connected. Thus, the sub-systems are not mutually exclusive. It facilitates the model to capture dependencies across the different groups in the market. The structure of the market graph, and thereafter the sub-systems are highly dependent on the threshold value. A very low  threshold would produce a complete graph and therefore would reduce the sampling procedure to random selection. Thus, the original LSSA becomes a special case of the proposed method under low threshold conditions. A very high threshold would produce an unconnected graph, with single assets in some sub-systems. Accordingly, the right balance of the threshold is needed to ascertain that the resulting market graph is neither too sparse nor too dense.

A suitable trade-off between the sub-system size $N_g$ and the number of samples $N_s$ is also needed. More samples do not necessarily mean better results, although larger sub-systems do indicate improved performance, the latter being dependent on the quantum hardware capability. The MIS-based sampling procedure presented here creates far fewer samples (worst-case $N$ samples) without compromising the quality of results. It builds problem-specific sub-systems that capture most of the strongest interactions between the variables in a limited number of samples. It not only allows large-scale portfolio optimization problems but also ensures excellent performance.

\begin{figure}[h!]
    \centering
    \includegraphics[scale=0.60]{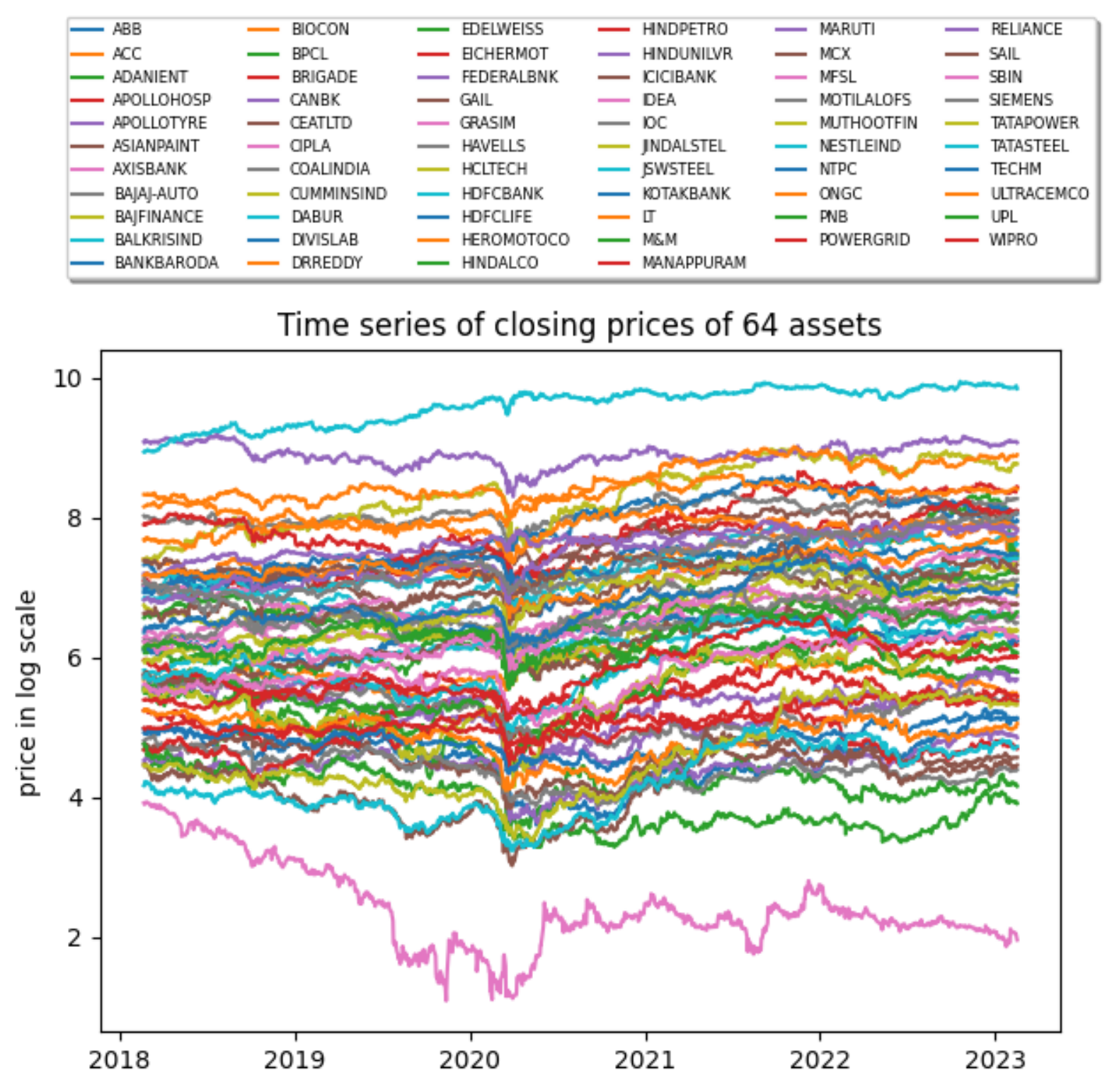}
    \caption{Closing prices of the 64 assets over the past 5 years shown in log scale}
    \label{fig:time_series}
\end{figure}

\subsubsection{Level 2 LSSA with MIS (random sampling)} \label{lvl2}

The method proposed in Section \ref{lvl1} requires a $N$ qubit machine to solve a portfolio optimization problem of $N$ assets. To accommodate much larger problems on the available hardware, we present a second-level division of the MIS problem itself. We use LSSA in its original form to first solve the MIS of the market graph and then use the final result to extract the placeholder assets, and subsequently create sub-systems for the portfolio optimization problem.  To elaborate further, the MIS problem is addressed as follows - a) random sampling to create sub-systems with fewer variables than the original problem, b) solving the smaller problems using quantum optimization methods, c) recombining the solutions using VQE to approximate the ground state of the original MIS problem.  Algorithm \ref{alg:method} shows the step-by-step procedure of solving a portfolio optimization problem of $N$ assets using the proposed methodology.

\begin{algorithm}
\setstretch{1.5}
\caption{MIS based portfolio optimization} \label{alg:method}
\begin{algorithmic}
\renewcommand{\algorithmicrequire}{\textbf{Input: }}
\renewcommand{\algorithmicensure}{\textbf{Output: }}
    \Require $\mu, \Sigma, \alpha, N, N_g, N_s $
    \Ensure Optimal asset investment for maximum returns and minimum risk
    \State Create a market graph $G = (V, E)$ of \textit{N} assets by thresholding the correlation matrix at $\alpha$;
    \State Create $N_s$ sub-systems each of size $N_g$ by random sampling; 
    \State $i \gets 1$
    \State $\texttt{results} \gets \begin{bmatrix} ... \end{bmatrix}_{N_s \times N}$
    \While{$i \leq N_s$}   
        \State $\texttt{results[i]} \gets x : \argmin\limits_{x} H = -\sum_{j \in V} x_j + 2 \sum_{(j,k) \in E} x_j x_k$
        \State $i \gets i + 1$
    \EndWhile
    \State $ \mathcal{S}^{wc} = \sum_{i=1}^{N_s} C^{(i)} \cdot \texttt{results[i]} $
    \State Apply VQE to optimize the coefficients $C^{(i)}$
    \State $\texttt{MIS\_assets} \gets sign(\mathcal{S}^{wc})$ \Comment{End of MIS} \\
    \State \texttt{portfolio\_subsystems} $ \gets [...]$
    \For{\texttt{asset} in \texttt{MIS\_assets}}
        \State \texttt{sub\_system} $\gets N_g$ highest correlated vertices to \texttt{asset} in $G$;
        \State add \texttt{sub\_system} to \texttt{portfolio\_subsystems}
    \EndFor
    \State $i \gets 1$
    \State $k \gets count(\texttt{MIS\_assets})$
    \State $\texttt{portfolio\_results} \gets \begin{bmatrix} ... \end{bmatrix}_{k \times N}$
    \While{$i \leq k$}   
        \State $\texttt{portfolio\_results[i]} \gets x : \argmin\limits_{x} H = - \mu^{T}x + \gamma x^{T}\Sigma x  $
        \State $i \gets i + 1$
    \EndWhile
    \State $ \mathcal{S}^{wc} = \sum_{i=1}^{k} C^{(i)} \cdot \texttt{portfolio\_results[i]} $
    \State Apply VQE to optimize the coefficients $C^{(i)}$
    \State $\texttt{portfolio\_assets} \gets sign(\mathcal{S}^{wc})$ \Comment{End of Portfolio Optimization}
\end{algorithmic}
\end{algorithm}

Unlike the conventional portfolio optimization problem, the MIS formulation in \eqref{mis_qubo} shows that quadratic interactions between all the variables are equally valued. This means that any selection of assets for the sub-systems will be mathematically similar to any other in approximating the full system objective. It is, therefore, justified that random sampling to construct sub-systems is as good as any other method in this case. However, there is now a need for more samples. Yet, the number of samples $N_s$ is not substantial because the market graph is not usually dense. The density of the graph depends precisely on the threshold value, which can be chosen to control the number of edge connections as $\mathcal{O}(N)$. With a little tuning, a reasonable estimation can be drawn for both the threshold value and the number of samples. 

\section{Implementation} \label{implementation}

The proposed method is tested on real data from the Indian stock market \cite{YahooFin} with the D-Wave Advantage\_system4.1 \cite{Dwave} for sub-system size $N_g = N/2$ for $N \in \{8, 16, 32, 40, 64\}$ and $N_g = N/4$ for $N \in \{40, 64\}$. The number of samples drawn is limited to a maximum of $N/2$ after testing for higher values without significant improvement. The VQE amplitude optimization is carried out on a local shot-based Qiskit \cite{qiskit} simulator of $N_{gb} = \lceil log_2 N_s \rceil$ qubits with 2048 shots. We use data over the past 5 years to compute the expected returns and co-variances of the assets. The Indian stock market data for $N=64$ assets is shown in Fig. \ref{fig:time_series}. The closing price of the assets is shown in a \textit{log} scale for better visibility.  The risk-aversion factor $\gamma$ in \eqref{pf_form} is chosen as 0.5 in all the experiments. The classical optimizer to optimize the parameters of VQE is COBYLA. Experiments showed no substantial differences in using other optimizers. The threshold value, $\alpha$ for the construction of the market graph is taken as 0.25 (meaning all asset connections with absolute strength less than 0.25 are removed) after careful assessment of the sparsity of the resulting graph. The python library PyQUBO \cite{pyqubo, pyqubo2} was used to build the QUBO models. In this implementation, we limit to simulated and quantum annealing to solve the QUBOs of the individual sub-systems and use a local simulator to apply the VQE for amplitude estimation. 

\begin{figure}[h!]
    \centering
    \begin{subfigure}[b]{0.49\textwidth}
        \includegraphics[width=\textwidth]{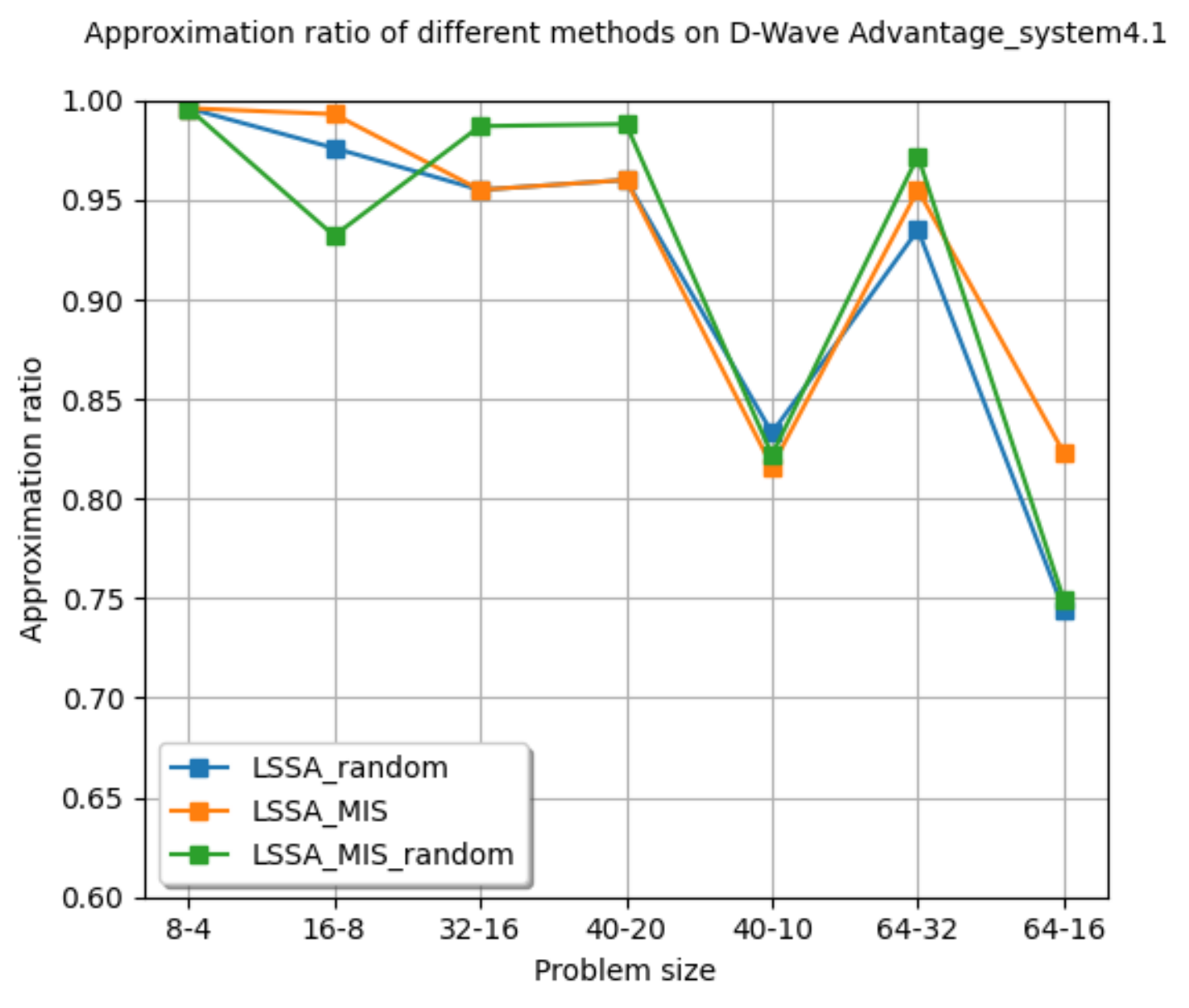}
        \caption{Experiments run on D-Wave Advantage\_system4.1}
        \label{fig:dwave_approx}
    \end{subfigure}
    \hfill
    \begin{subfigure}[b]{0.49\textwidth}
        \includegraphics[width=\textwidth]{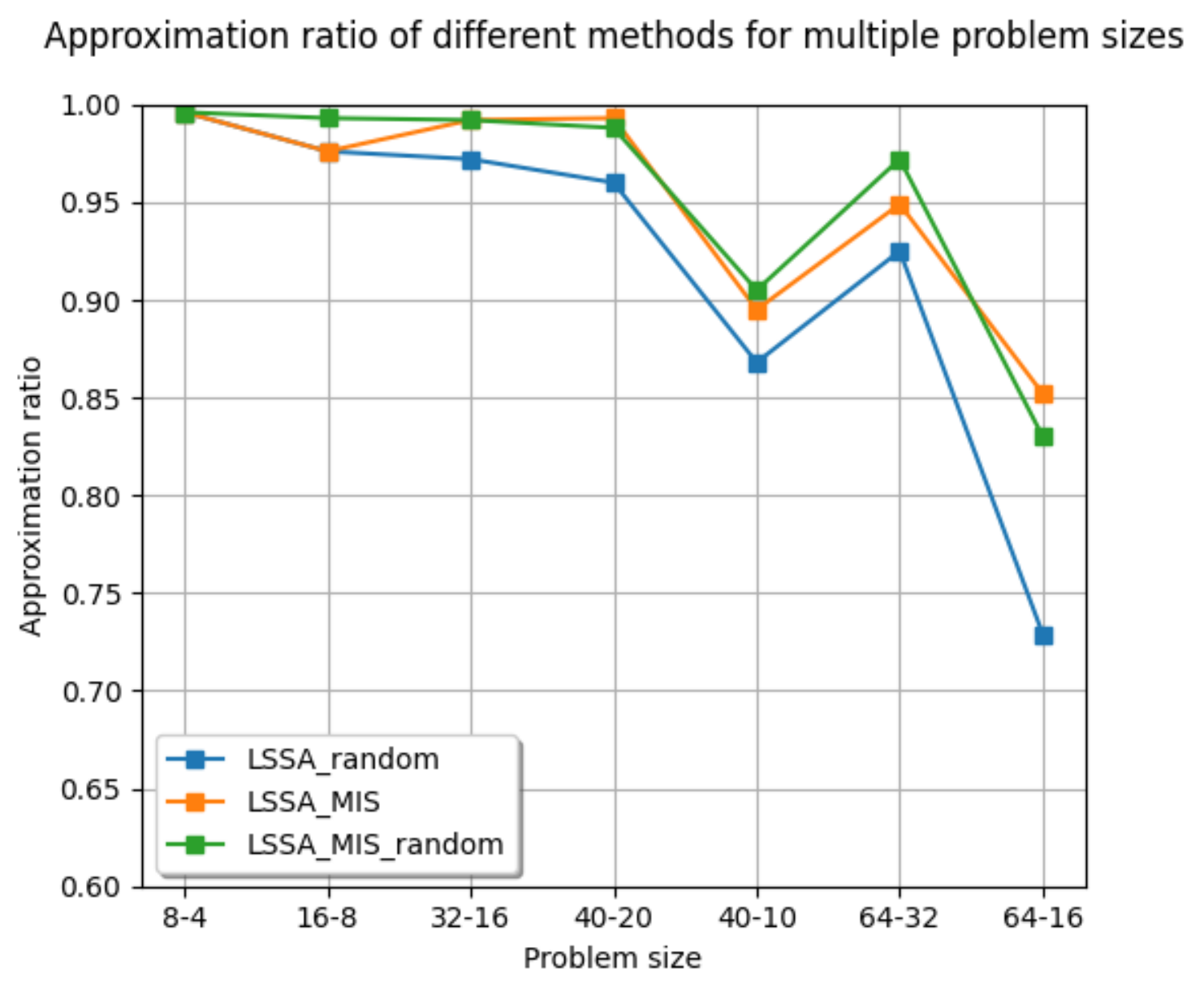}
        \caption{Experiments run on simulated annealing}
        \label{fig:sim_approx}
    \end{subfigure}
        \caption{Approximation ratios of different methods. The labels indicate the different techniques used to solve the problem. \texttt{LSSA\_random} represents the original LSSA approach with random sampling, \texttt{LSSA\_MIS} represents the approach described in Section \ref{lvl1} and \texttt{LSSA\_MIS\_random} is the process explained in Section \ref{lvl2}}
        \label{fig:approx_ratios}
\end{figure}

\section{Results and Discussion} \label{results_and_discussion}

To establish a comparison of performance between the different methods, we define an approximation ratio 
\begin{equation}
    R_{ar} = \frac{\texttt{Method GSE}}{\texttt{Classical GSE}} 
\end{equation}
where GSE is the Ground State Energy, Method refers to the different methods, and Classical GSE is the Exact solution energy for $N < 20$ and \texttt{D-Wave-Tabu} solver energy for $N \geq 20$ assets. Overall, an $N$ asset portfolio optimization problem is solved by first deriving $N_s$ samples and solving  MIS of the market graph on a $N_g$ qubit quantum annealer and $\lceil log_2 N_s \rceil$ qubit gate-based machine. The results are then used to solve the original portfolio problem on a $N_g$ qubit quantum annealer and $\lceil log_2 |MIS| \rceil$ qubit gate-based machine ($ |MIS| \ll N$ is the cardinality of the Maximum independent set). The total calls made to the annealer are $N_s + |MIS|$.

\begin{figure}[h!]
    \centering
    \begin{subfigure}[b]{0.49\textwidth}
        \includegraphics[width=\textwidth]{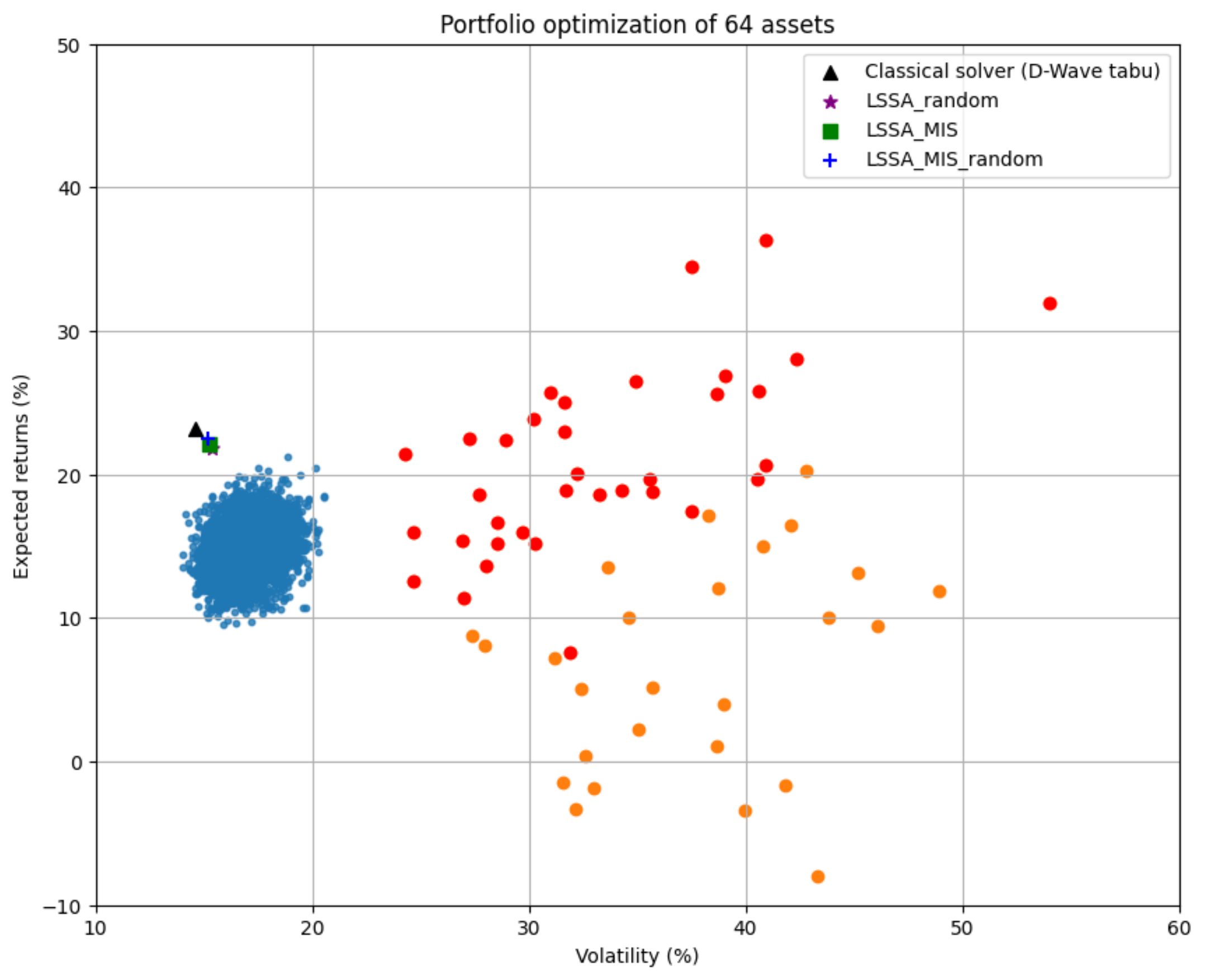}
        \caption{}
        \label{fig:efficient_frontier_orig}
    \end{subfigure}
    \hfill
    \begin{subfigure}[b]{0.49\textwidth}
        \includegraphics[width=\textwidth]{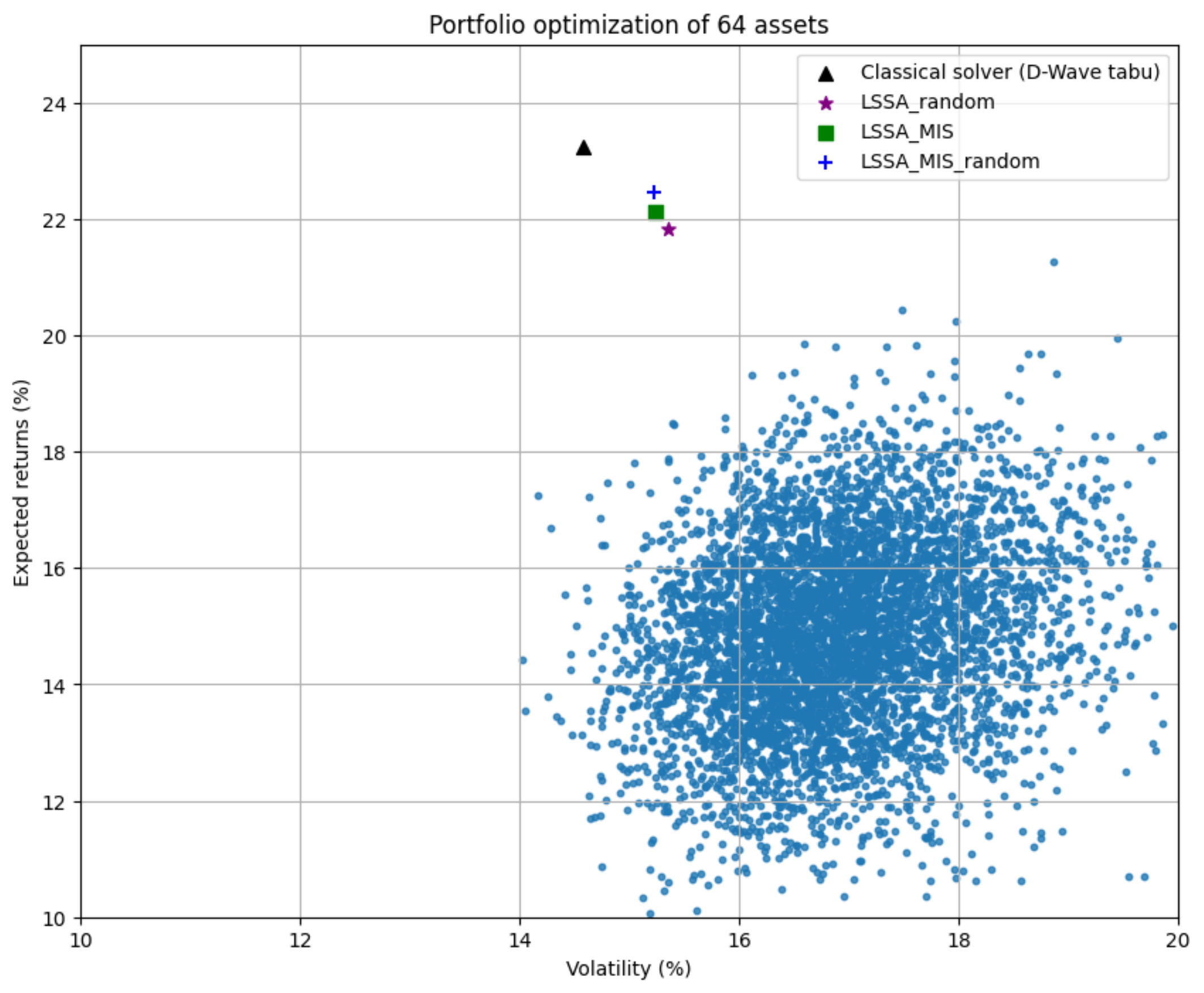}
        \caption{}
        \label{fig:efficient_frontier_zoom}
    \end{subfigure}
        \caption{Return and volatility of the portfolio optimization problem with $N = 64$ assets. In Fig. (a), the scattered dots in blue represent results from a random sampling of 5000 different portfolios. The orange and red dots represent individual assets and the stocks chosen by \texttt{LSSA\_MIS\_random}, respectively. The axes are clipped for better visibility. In Fig. (b), the results obtained from classical solver \texttt{D-Wave Tabu} are shown by the triangle, the \texttt{LSSA\_random} by the star, the \texttt{LSSA\_MIS} by the square and \texttt{LSSA\_MIS\_random} by the plus. Fig. (b) is a zoomed-in version of Fig. (a)}
        \label{fig:efficient_frontier}
\end{figure}

The approximation ratios are shown in Fig \ref{fig:approx_ratios}. On the horizontal axis, the ticks represent the two values $N$ (\textit{problem size}) and $N_g$ (\textit{sub-system size}) separated by a hyphen ($-$). So, 40-20 corresponds to a portfolio optimization problem of 40 assets solved by creating samples each of size 20. The ratios shown here are derived by experimenting with different values of $N_s$ and choosing the one that is best among them. The results indicate that the proposed method performs at par with the Tabu solver with much fewer samples than the original method. Specifically, for $N = 64$ assets and $N_g = 32$, \texttt{LSSA\_random} requires $N_s = 32$ samples, while \texttt{LSSA\_MIS} and \texttt{LSSA\_MIS\_random} use only $N_s = 12$ and $N_s = 13$ samples respectively. Table \ref{tbl: results_data} shows the experimental data for the least number of sub-systems required by different methods for the performance indicated in Fig. \ref{fig:approx_ratios}.

\begin{table}
\centering
\caption{Experimental data showing the proposed method using fewer sub-systems. \texttt{MIS} refers to the number of samples derived for solving MIS problem and \texttt{PO} refers to the number of samples used for Portfolio Optimization problem.}
\vspace{3mm}
\label{tbl: results_data}
\resizebox{0.65\columnwidth}{!}{%
\begin{tabular}{@{}ccccc@{}}
\toprule
$N$                    & $N_g$                & \multicolumn{3}{c}{$N_s$}                                             \\ \midrule
\multicolumn{1}{l}{} & \multicolumn{1}{l}{} & \multicolumn{1}{l}{\texttt{LSSA\_random}} & \multicolumn{1}{l}{\texttt{LSSA\_MIS}} & \multicolumn{1}{l}{\texttt{LSSA\_MIS\_random}} \\
8                    & 4                    & 4                     & 4                     & 4 - \texttt{MIS}, 4 - \texttt{PO}                \\
16                   & 8                    & 4                     & 4                     & 4 - \texttt{MIS}, 4 - \texttt{PO}                 \\
32                   & 16                   & 16                    & 8                     & 8 - \texttt{MIS}, 8 - \texttt{PO}                 \\
40                   & 20                   & 16                    & 7                     & 8 - \texttt{MIS}, 8 - \texttt{PO}                 \\
40                   & 10                   & 8                     & 8                     & 8 - \texttt{MIS}, 8 - \texttt{PO}              \\
64                   & 32                   & 32                    & 12                    & 16 - \texttt{MIS}, 13 - \texttt{PO}           \\
64                   & 16                   & 16                    & 16                    & 16 - \texttt{MIS}, 16 - \texttt{PO}            \\ \bottomrule
\end{tabular}%
}
\end{table}

Fig. \ref{fig:efficient_frontier} shows the efficient frontier and represents the expected returns at various levels of volatility (risk) for a problem with 64 assets and a sub-system size of 32 variables. The individual assets are shown in orange dots (for better visualization, the axes are clipped). The assets chosen for investment by the \texttt{LSSA\_MIS\_random} procedure are shown in red. The return and volatility of the portfolio formed by choosing these assets are shown in a blue plus symbol which is substantially better than random picking of stocks. To put things into perspective, random combinations of assets are picked and their corresponding values are shown as scattered dots in blue. The proposed method only uses 13 samples for portfolio optimization and 16 samples for solving MIS, while the baseline \texttt{LSSA\_random} uses 32 samples.

Although the results shown here are stochastic in nature, the data from a number of experiments indicate that the proposed method matches the baseline method in terms of results and thus provides a practical and feasible approach to portfolio optimization using quantum computing and quantum annealing.

\section{Conclusion and Future Work} \label{conlusion}

In this paper, we introduced an efficient way to create sub-systems for the LSSA technique. We demonstrated our algorithm on a simple portfolio optimization problem on a real dataset. Our approach significantly reduces the number of samples required compared to the earlier method, i.e. it lowers the requirements to apply the technique on real quantum hardware for practically relevant problems. We recognise that the MIS-based method is applicable when the assets are from different sectors of the market and thus have different degrees of correlation in general. LSSA becomes a specific case of the proposed method for a small set of strongly correlated assets. Therefore, the proposed method is best suited for problem instances where there are grades of diversity, which is usually true in a real setting.

In this study, we have relied on solving the sub-systems using Quantum Annealing. However, even gate-based quantum chips can be used. Algorithms like the Grover Adaptive Search \cite{gas} could be better methods to solve the individual sub-systems. The VQE step to estimate the coefficients of the ground state combinations might be replaced with the Variational non-orthogonal optimization strategy \cite{variational_non_orthogonal} to allow much larger problem sizes. We aim to explore these ideas in the future. There are still several practical considerations to the problem of portfolio optimization. Besides being able to solve large-scale problems efficiently, the solutions must adhere to practical constraints and be flexible to accommodate changes as per market conditions. Overall, our findings suggest that a hybrid of annealing and gate-based quantum computing can be a promising tool for portfolio optimization, and we look forward for further exploration of this exciting field. 

\section*{Acknowledgements}

The authors express their sincere gratitude to Mr. Ankit Khandelwal, Mr. Manoj Nambiar and Dr. Gautam Shroff from TCS Research for their invaluable insights, unwavering support and encouragement.

%Bibliography
\bibliographystyle{unsrt}  
\bibliography{references}

\end{document}